\documentclass[twocolumn,prb,aps]{revtex4}
\usepackage{graphicx}
\usepackage{epsfig}
\begin{document}
\title{\Large Tuning of the superconducting and ferromagnetic transitions by Cu doping
 for Ru in GdSr$_2$RuCu$_2$O$_8$}
\author{\Large J. Janaki$^1$, T. Geetha Kumary$^1$, R. Nagarajan$^2$, T. A. Mary$^3$,
M. C. Valsakumar$^1$, V. S. Sastry$^1$, Y. Hariharan$^1$ and T. S. Radhakrishnan$^1$}
\affiliation{\large $^1$ Materials Science Division, IGCAR Kalpakkam 603 102, India\\
$^2$ Department of CMP\&MS, TIFR, Mumbai 400 005, India\\
$^3$ Materials Technology Division, IGCAR Kalpakkam 603 102, India}
\vskip 1cm
\begin{abstract}
\noindent{\bf Abstract}\\[5mm]
In order to explore the possibility of tuning  superconducting and ferromagnetic
 transitions by Cu doping (for Ru) in GdSr$_2$RuCu$_2$O$_8$, we have carried out
 synthesis and characterization of  GdSr$_2$Ru$_{1-x}$Cu$_{2+x}$O$_8$ (x = 0, 0.05, 0.1, 0.2)
 and studied their physical properties. Coexistence of superconductivity and
 ferromagnetism is observed in all the Cu doped samples studied here. The zero
 field susceptibility data suggests formation of a spontaneous vortex phase. Cu
 doping decreases the ferromagnetic Curie temperature, whereas the
 superconducting transition temperature increases until an optimal concentration
 x $\sim$0.1. This reflects an increase in hole transfer to the CuO$_2$ planes and
 reduction of ferromagnetic order within the ruthenate layers.\\[5mm]
\end{abstract}
\maketitle
\vskip 1cm
\section{Introduction}

Superconductivity (SC) and ferromagnetism (FM) are generally believed to be two
 mutually antagonistic  orders. However, many interesting consequences like
 coexistence of triplet SC with FM, formation of a spontaneous vortex phase \cite{ref1},
 spatially inhomogeneous Fulde-Ferrel-Larkin-Ovchinnikov type superconducting
 order \cite{ref2}, etc. can arise if such a coexistence occurs. Thus, the observation
 of SC (with a high superconducting transition temperature T$_c$  $\sim$ 46K) coexisting
 with FM (with the magnetic transition temperature T$_m$ $\sim$ 132K) in the hybrid
 rutheno-cuprate  GdSr$_2$RuCu$_2$O$_8$ \cite{ref3,ref4,ref5} has motivated  intense activity to
 characterize the families of compounds \cite{ref6} LnSr$_2$RuCu$_2$O$_8$ and
 Ln$_{1+x}$Ce$_{1-x}$Sr$_2$RuCu$_2$O$_{10}$ (Ln = Sm, Eu, Gd), and to study the nature of
 superconductivity, magnetism and their inter-relation.

GdSr$_2$RuCu$_2$O$_8$ is isostructural with the well known YBa$_2$Cu$_3$O$_{7-\delta}$ with Y and Ba
 being completely replaced by Gd and Sr respectively, and the CuO chain replaced
 by RuO$_2$ planes. While the current broad understanding of the system is that SC
 originates in the CuO$_2$ planes and FM in the RuO$_2$ planes, details of the nature
 of both SC and magnetism require further understanding. GdSr$_2$RuCu$_2$O$_8$ has a very
 low H$_{c1}$ ($<$  $\sim$100 Oe) and the superconducting properties are highly dependent on
 the details of sample preparation. Rietveld refinement of the x-ray diffraction
 pattern of GdSr$_2$RuCu$_2$O$_8$ showed \cite{ref7} that oxygen annealing leads to variation of
 the cation composition without  change in the oxygen content. The formal charge
 of Cu, which is less than the optimal value (for the occurrence of SC) for the
 as-prepared sample, increases upon prolonged annealing \cite{ref7}.  It thus appears
 that the purity of the starting materials and the heating schedules play a
 crucial role in the structure and microstructure of the final product \cite{ref8},
 thereby rationalizing contradictory reports \cite{ref9} of existence and non-existence
 of SC in this compound.
\begin{figure*}[t]
\centerline{\epsfig{figure=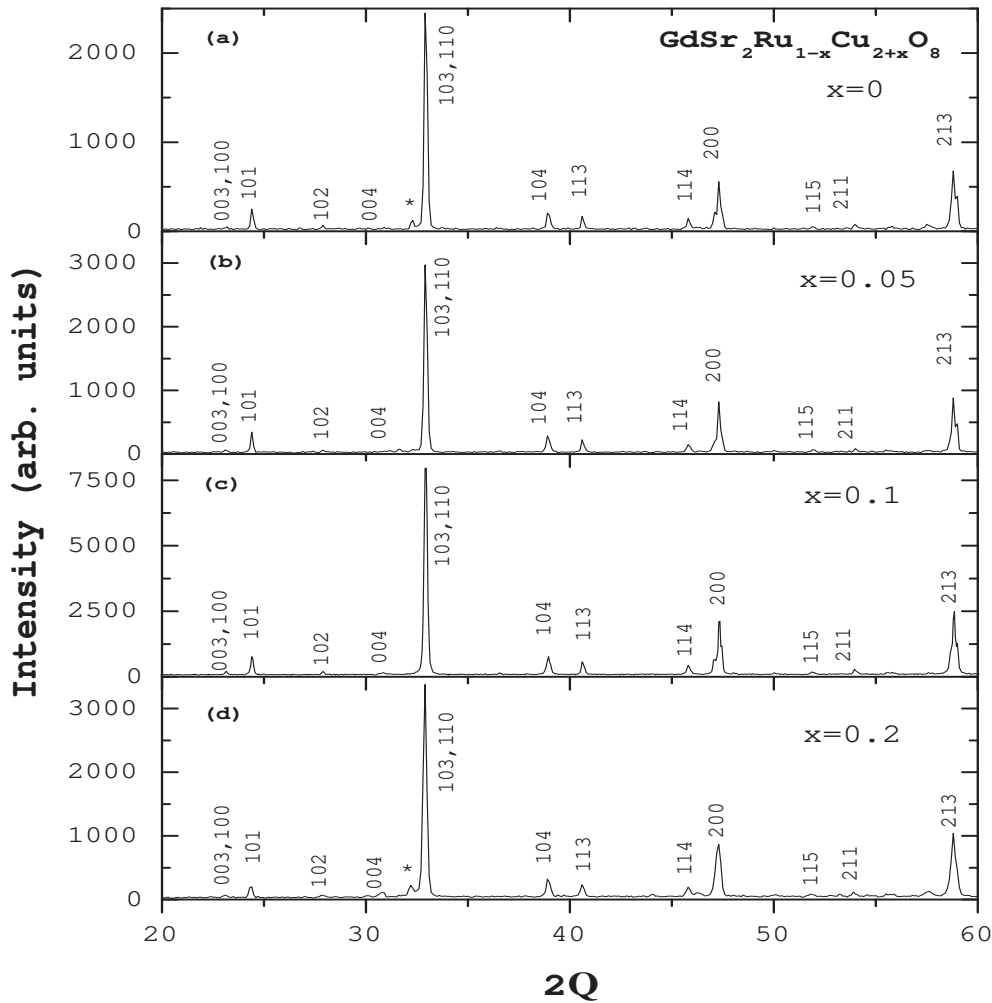,height=6in,width=6in}}
\caption{X-ray diffraction patterns of GdSr$_2$Ru$_{1-x}$Cu$_{2+x}$O$_8$ for (a) x = 0, (b) x=0.05,
 (c) x=0.1 and (d) x=0.2. The lines corresponding to the impurity phases are
 shown with *.}
\end{figure*}

The neutron diffraction studies \cite{ref10,ref11,ref12} suggest that the magnetic order in
 GdSr$_2$RuCu$_2$O$_8$ is predominantly G-type antiferromagntic alignment of the Ru
 moments with a low temperature moment $\sim$1 mB along c axis. However, the
 magnetization studies \cite{ref11,ref13} show a spontaneous magnetization Ms $\sim$0.1mB  (per
 Ru). These two results can be reconciled if the Ru moments are canted to give a
 net magnetic moment in a direction perpendicular to the c axis. Neutron
 diffraction study \cite{ref14} of the related compound YSr$_2$RuCu$_2$O$_8$ showed a spontaneous
 magnetic moment $\sim$0.28 mB (which could be enhanced by applying an external
 magnetic field) perpendicular to the c axis. A recent analysis \cite{ref15} of the
 magnetization data suggests I-type ordering consistent with a band structure
 calculation \cite{ref16}, but inconsistent with the neutron diffraction studies.
 Further, the magnetism will also be affected by the valence of Ru. The formal
 valence of Ru has been taken to be 5+ in most of the previous Rietveld
 analyses. Recent NMR \cite{ref17} and XANES \cite{ref18} studies indicate mixed valence for Ru
 with 40\% Ru$^{4+}$  and 60\% Ru$^{5+}$ in  GdSr$_2$RuCu$_2$O$_8$. This also entails several
 possibilities \cite{ref18} for the magnetic order in this compound.  Thus, the precise
 nature of magnetic order in this compound still remains controversial and needs
 to be explored.

The effect of substitution on the physical properties of this system is also
 very interesting and has been reported in literature \cite{ref3,ref4,ref19,ref21}. Sn doping
 \cite{ref19} suppresses T$_m$ from 138K for x = 0 to 78K for x = 0.4 in GdSr$_2$Ru$_{1-x}$Sn$_x$Cu$_2$O$_8$
 and T$_c$ increases from 36K at x = 0 to 48K at x = 0.2 but falls thereafter. On
 the other hand even a small fraction of Zn (for Cu) rapidly reduces T$_c$ and
 superconductivity is completely suppressed for the 3\% substituted sample \cite{ref3,ref4}.
 Motivated by above considerations, we have undertaken a study of tuning of the
 superconducting and magnetic properties by Cu doping for Ru in GdSr$_2$RuCu$_2$O$_8$. In
 this regard, and in continuation of our previous work \cite{ref20}, we have prepared
 GdSr$_2$Ru$_{1-x}$Cu$_{2+x}$O$_8$ (x = 0, 0.05, 0.1, 0.2) and investigated their properties,
 which are presented below.

\begin{figure*}
\centerline{\epsfig{figure=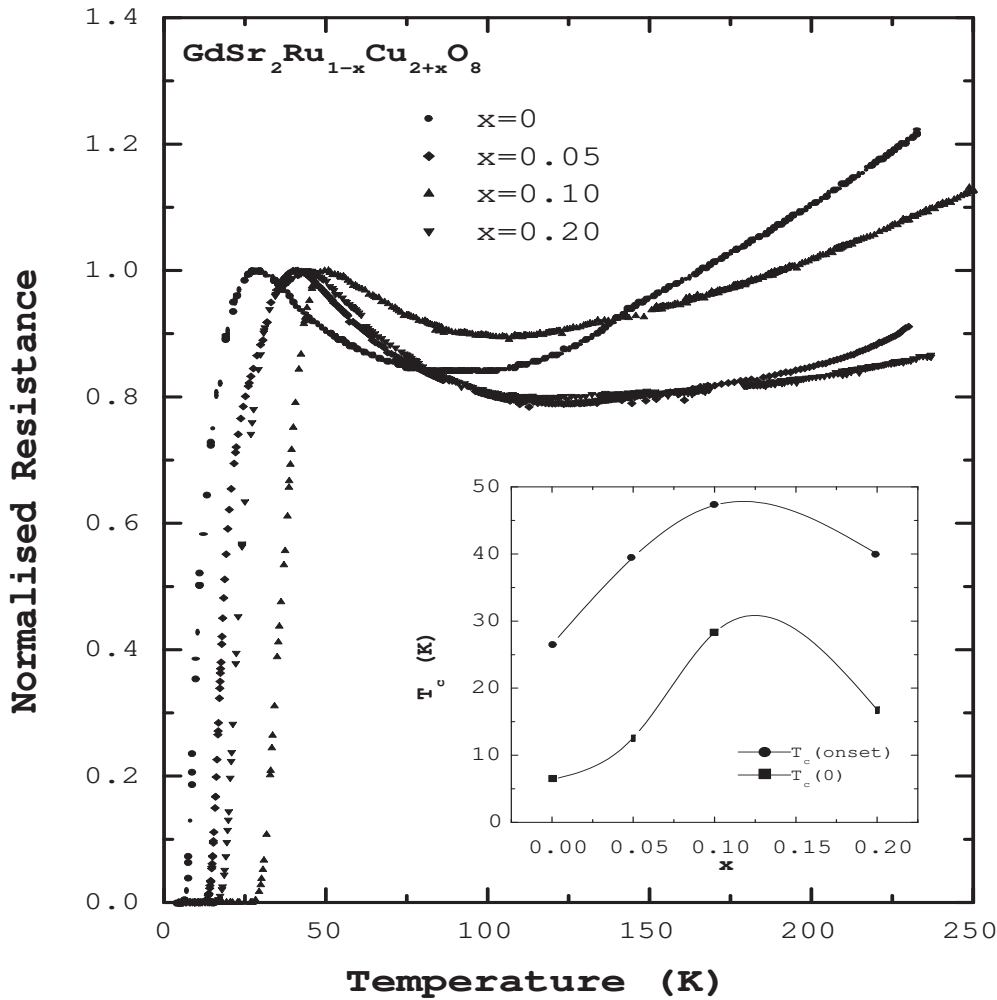,height=6in,width=6in}}
\caption{Normalised Resistance vs temperature curves of GdSr$_2$Ru$_{1-x}$Cu$_{2+x}$O$_8$ for x=0,
 0.05, 0.1 and 0.2.  T$_c$ corresponding  to  onset  and  zero  resistance  are
 shown  as  a  function  of  x  in  the  inset.}
\end{figure*}

\section{Synthesis and Characterization}

The compounds have been synthesized by the high temperature solid state reaction
 technique, which is commonly employed for the synthesis of high temperature
 superconductors. Stoichiometric quantities of Gd$_2$O$_3$, SrCO$_3$, CuO and Ru powder,
 all $>$  99.99\% pure, have been mixed and ground. The mixture (in the form of
 powder) is heated at 600$^{\circ}$C for 48 hours to avoid RuO$_2$ volatility and then at
 950$^{\circ}$C, in air, for 24 hours. It is then compressed into pellets and then heated
 at 1000$^{\circ}$C in air for 72 hours. Repeated grinding and heating at 1000$^{\circ}$C in air
 is necessary to obtain phase pure samples. A final step involving annealing at
 1060$^{\circ}$C in oxygen for 96 hours is necessary to obtain the desired
 superconducting characteristics. High resolution X-ray powder diffraction
 analysis is carried out using a STOE Powder diffractometer in the step scanning
 mode. Electrical resistivity (four probe method, 4.2-300K) and AC
 Susceptibility (4.2-300K) have been carried out using home-assembled
 equipments. D.C. Magnetization studies have been performed using a Quantum
 Design SQUID Magnetometer (5K-300K). All the samples have been prepared in two
 batches to test for reproducibility.

\section{Results and Discussions}
\begin{figure*}
\centerline{\epsfig{figure=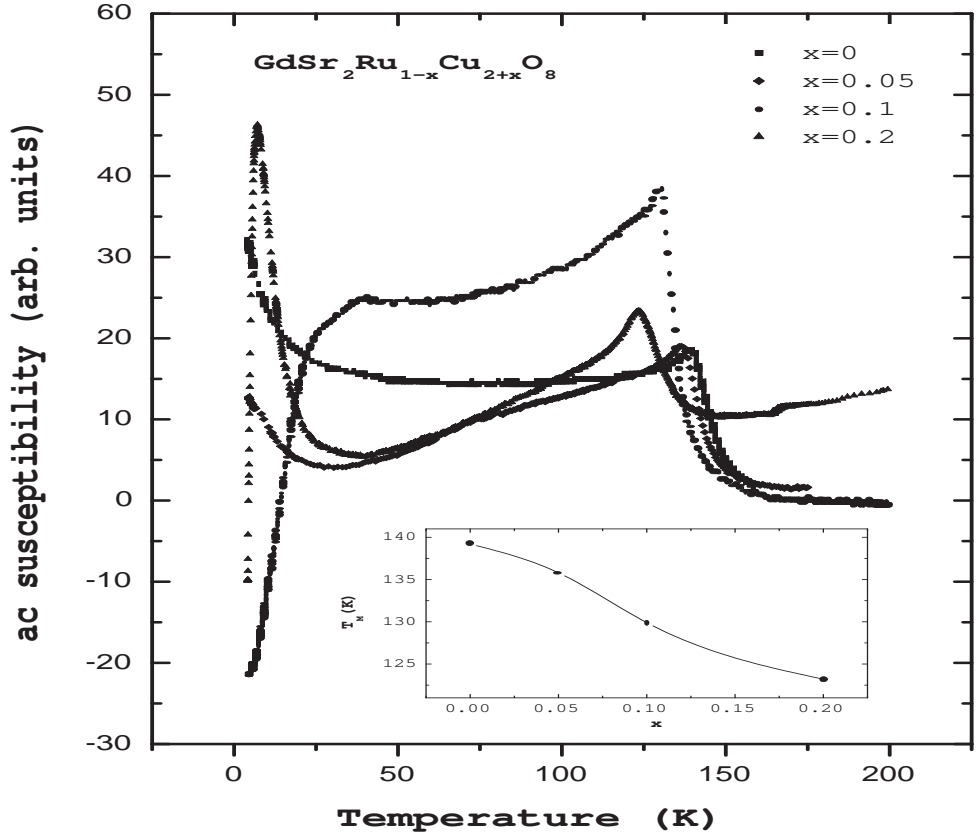,height=5in,width=6in}}
\caption{ AC susceptibility vs temperature curve for GdSr$_2$Ru$_{1-x}$Cu$_{2+x}$O$_8$ for x=0, 0.05,
0.1 and 0.2. Magnetic transition  temperatute  T$_m$  is  shown  as  a  function
 of  x  in  the  inset.}
\end{figure*}
Figure 1a shows the XRD of GdSr$_2$RuCu$_2$O$_8$. Phase analysis by XRD indicates that
 the impurity phase content is $<$   5\%. The XRD pattern has been analyzed on the
 basis of the reported tetragonal structure with space group P4/mmm, and lattice
 constants have been found to be a=3.83$\AA$ and c=11.58$\AA$ which are in close
 agreement with the values reported in the literature \cite{ref3,ref4}. From the analysis
 of several other samples of GdSr$_2$RuCu$_2$O$_8$ it is estimated that the detectability
 limit of the impurity phases like SrRuO$_3$ and GdSr$_2$RuO$_6$ type phases and CuO by
 XRD is $<$  1\%. The crystallite sizes of the sample are small $\sim$500-1000$\AA$ as
 calculated from XRD line widths using the Scherrer formula. Figure 2 shows the
 resistance (R) vs temperature (T) curve of GdSr$_2$Ru$_{1-x}$Cu$_{2+x}$O$_8$. This plot clearly
 indicates the occurrence of superconductivity in these samples, with T$_c$ onset
 of 26.5K and zero resistance at around 6.4K for the stoichiometric composition
 GdSr$_2$RuCu$_2$O$_8$. The broadness of the transition may be due to the small
 crystallite size. Figure 3 shows the ac susceptibility, $\chi$, of GdSr$_2$Ru$_{1-x}$Cu$_{2+x}$O$_8$
 as a function of temperature. This figure shows peaks corresponding to the
 magnetic ordering of the samples with the peak occurring at 139.2K for the
 stoichiometric composition GdSr$_2$RuCu$_2$O$_8$. The  relatively small value of the
 signal indicates the weak nature of ferromagnetism of the material. On closer
 examination of the R vs T plot, the magnetic ordering transition can be seen as
 a change in slope at  $\sim$139.2K. The diamagnetic shift corresponding to the
 superconducting transition at 26.5K is not observed in the susceptibility data
 for the stoichiometric composition GdSr$_2$RuCu$_2$O$_8$ (Figure 3) below 26.5K. On the
 other hand, the $\chi$ increases with decrease in T below 26.5K (Figure 3). The
 steep increase in $\chi$ at low temperature arises due to the paramagnetic
 contribution of the Gd moments which order antiferromagneticaly at 2.5K. It is
 believed that the superconductivity is weak and is also perhaps not homogeneous
 with the result that the magnetic signal due to the paramagnetic contribution
 of the Gd moments overwhelms the diamagnetic superconducting signal. In order
 to confirm ferromagnetism in GdSr$_2$RuCu$_2$O$_8$, isothermal magnetization studies
 have been carried out as a function of field at T=5K. Figure 4a presents the
 M-H plot for GdSr$_2$RuCu$_2$O$_8$ along with an inset indicating the low field part of
 the hysteresis loop .The shape of the M-H curve (Fig. 4a) is characteristic of
 FM and shows weak hysteresis with a low coercive field of $\sim$200 Oe (Fig. 4a
 inset). The magnetization studies at high fields  indicate a saturation
 magnetization of  7 mB corresponding to ordering of Gd moments in high field.

Fig. 1 b, c and d show the XRD patterns of compounds GdSr$_2$Ru$_{1-x}$Cu$_{2+x}$O$_8$ (x =
 0.05, 0.1, 0.2). The samples with x = 0.05, x=0.1 and x=0.2 have impurity phase
 content approximately $<$  2\%,  $<$  1\% and $<$  3\%, respectively. All these compositions
 have the tetragonal structure with space group P4/mmm, isostructural with the
 stoichiometric composition. The lattice parameters of the compounds
 GdSr$_2$Ru$_{1-x}$Cu$_{2+x}$O$_8$ (x = 0.05, 0.1, 0.2,) are almost identical, a $\sim$ 3.83 $\AA$ and
 c $\sim$ 11.57 $\AA$ and do not differ significantly from the stoichiometric GdSr$_2$RuCu$_2$O$_8$.

\begin{figure*}
\centerline{\epsfig{figure=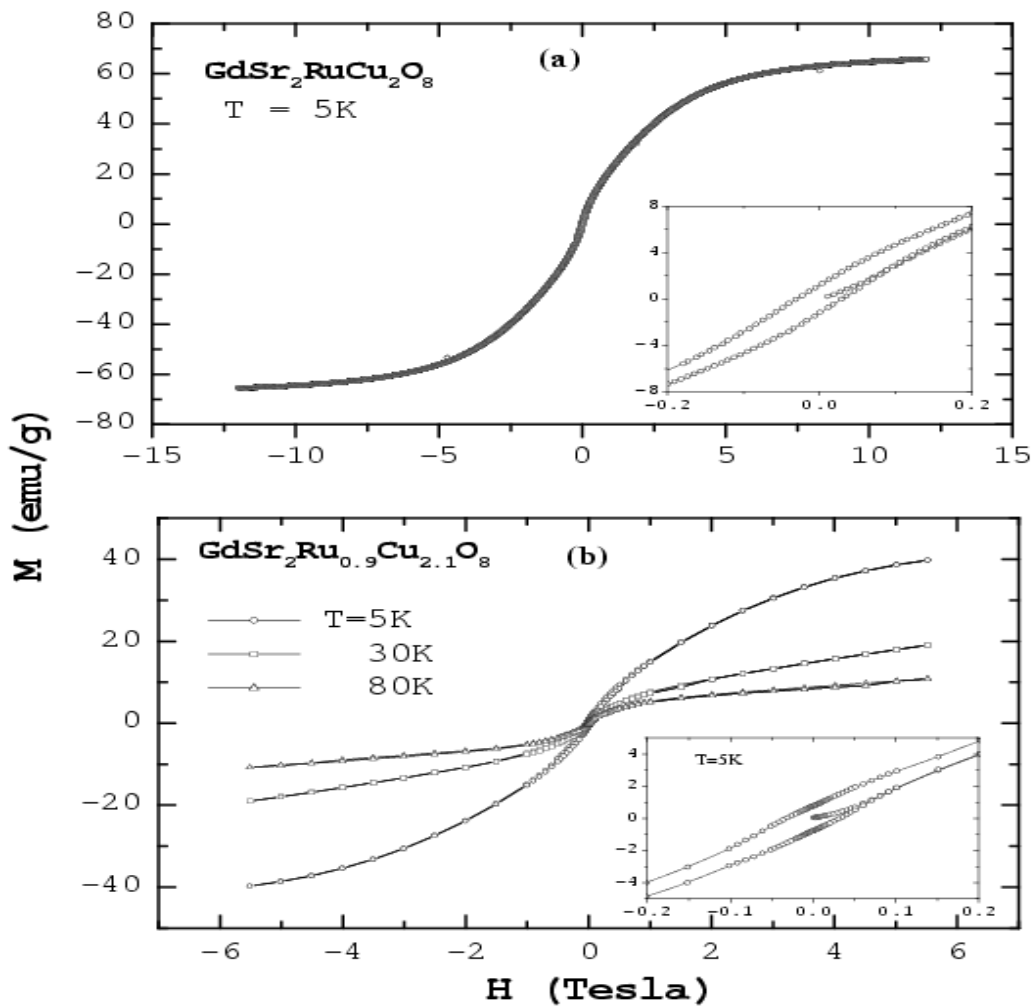,height=6in,width=6in}}
\caption{M vs H curve for (a) GdSr$_2$RuCu$_2$O$_8$  at 5K and  (b) GdSr$_2$Ru$_{0.9}$Cu$_{2.1}$O$_8$  at T=5K,
 30K and 80K. Inset       in  each  figure  shows  the  low  field  part  of the
  hysteresis  loop  at 5K.}
\end{figure*}

Figure 2 shows the temperature dependence of resistance for GdSr$_2$Ru$_{1-x}$Cu$_{2+x}$O$_8$ (x
 = 0.05, 0.1, 0.2) along with that of the x=0 sample. All of them exhibit
 superconductivity with zero resistance. The onset and zero resistance
 temperatures are   (26.5, 6.4), (39.5, 12.6), (47.2, 28.3) and (40.0, 16.8)K
 respectively for x = 0, 0.05, 0.1 and 0.2. One of the most important
 observations is that by Cu doping T$_c$ can be increased and x=0.1 appears to be
 the optimum Cu doping concentration.  It is also to be noted that these
 compounds are metallic over most of the temperature range down to 100K. Below
 100K there is a semiconducting-like upturn just above the transition to
 superconducting state. This may be due to the effect of granularity or magnetic
 scattering. Further experiments in a magnetic field would clarify this and are
 in progress. However it is interesting to note that the sample x=0.1, which
 shows the highest T$_c$, also has the minimum semiconducting upturn. The ac
 susceptibility vs temperature plots for  GdSr$_2$Ru$_{1-x}$Cu$_{2+x}$O$_8$ (x=0, 0.05, 0.1 and
 0.2) are shown in Figure 3. It may be noted that the undoped composition did
 not show a diamagnetic signal corresponding to superconductivity, whereas all
 the Cu substituted compositions show signature of superconductivity in the ac
 susceptibility measurements as well. This may be due to an increase in grain
 size and/or homogeneity. The ferromagnetic transitions are also observed in the
 Cu doped samples (Figure 3) but at a significantly lower temperature (the
 magnetic transition temperature T$_m$ being 135.8K, 129.9K and 123.2K respectively
 for x = 0.05, 0.1 and x=0.2  ) as compared to 139.2K for the stoichiometric
 GdSr$_2$RuCu$_2$O$_8$. Figure 4b shows the isothermal magnetization hysteresis at 5, 30
 and 80K for a typical Cu doped sample with composition GdSr$_2$Ru$_{0.9}$Cu$_{2.1}$O$_8$. This
 confirms the presence of weak ferromagnetism in the Cu doped samples similar to
 the stoichiometric GdSr$_2$RuCu$_2$O$_8$.

\begin{figure}
\centerline{\epsfig{figure=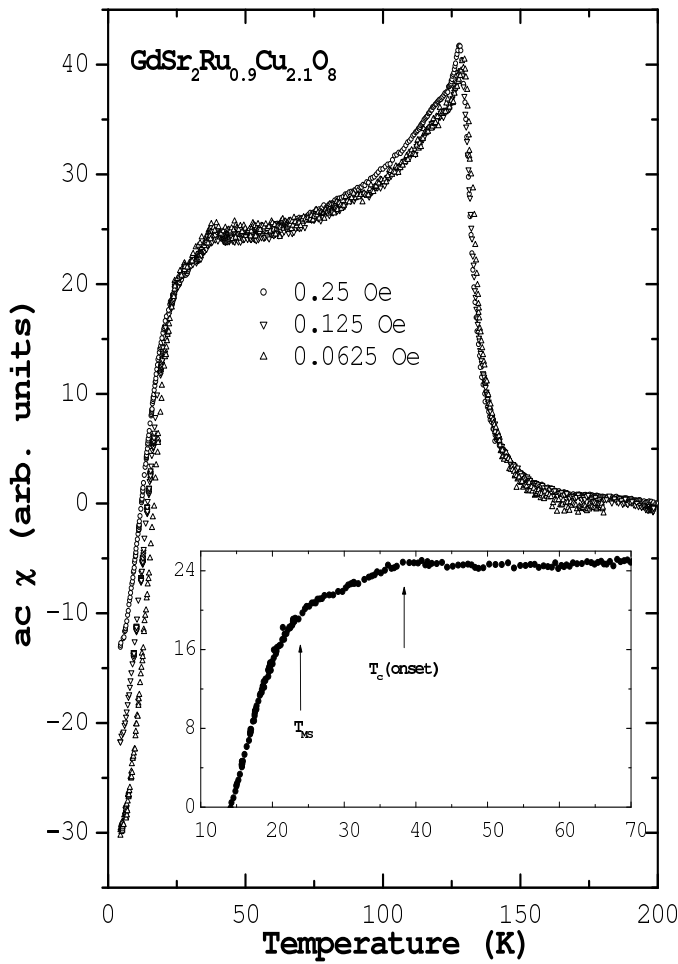,height=5in,width=3.5in}}
\caption{AC  susceptibility of GdSr$_2$Ru$_{0.9}$Cu$_{2.1}$O$_8$  for three different measuring
 fields. The change of slope in c  at  a  temperature T$_m$S  less  than  T$_c$
 (onset)  is  shown  in  the  inset.}
\end{figure}

As GdSr$_2$Ru$_{0.9}$Cu$_{2.1}$O$_8$ had the highest T$_c$ and also happens to have the highest
 purity ($<$   1\% impurity phase content) more detailed measurements have been
 carried out on it. Zero Field Cooled (ZFC) and Field Cooled (FC) measurements
 of ac susceptibility yielded identical results.  Fig. 5 shows the field cooled
 ac susceptibility in different measuring fields (amplitude = 0.25, 0.125 and
 0.0625 Oe) for GdSr$_2$Ru$_{0.9}$Cu$_{2.1}$O$_8$. The ac $\chi$ starts to decrease at T$\sim$38K
 indicating the onset of superconductivity. A significant change in slope of $\chi$
 vs T is observed  at T$_m$ $\sim$23K below the superconducting transition. This is
 perhaps an evidence of the transition from a bulk Meissner state to a
 spontaneous vortex state similar to that observed by Bernhard et. al.22 in dc $\chi$
 measurements. A smaller amplitude of the measuring field results in a stronger
 diamagnetic signal (see Fig. 5). The incomplete shielding effect is due to the
 low value of the lower critical field H$_{c1}$ that has been reduced by impurity
 scattering or small grain size.

\begin{figure}
\centerline{\epsfig{figure=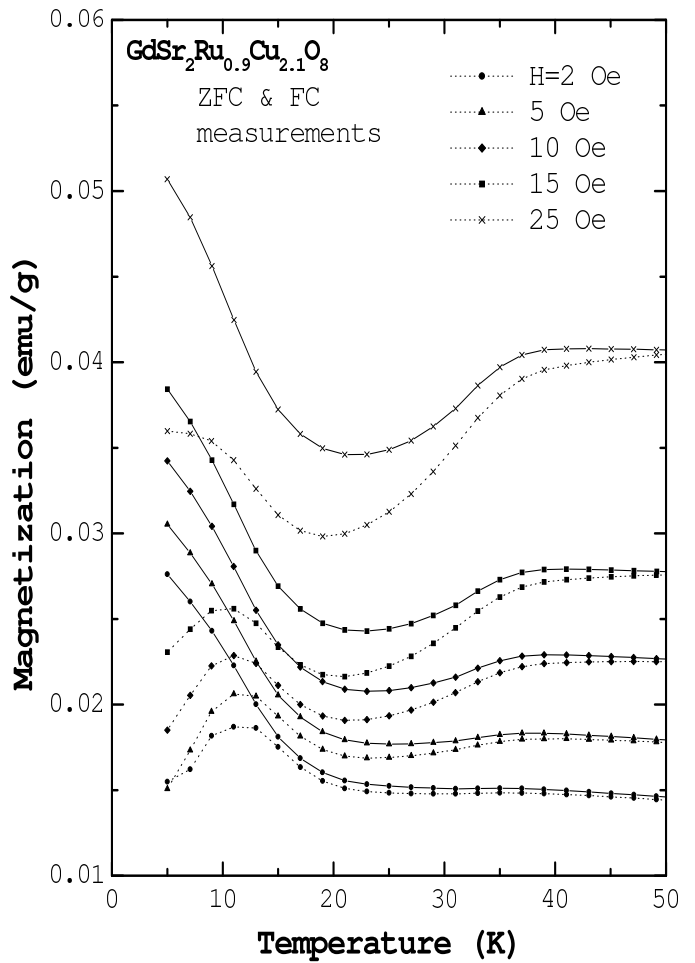,height=5in,width=3.5in}}
\caption{DC Magnetization vs Temperature of GdSr$_2$Ru$_{0.9}$Cu$_{2.1}$O$_8$ measured under
 different applied fields. Solid  and  dotted  lines  correspond  to Field
 Cooled and Zero Field Cooled measurements, respectively.}
\end{figure}

The dc magnetization studies on GdSr$_2$Ru$_{0.9}$Cu$_{2.1}$O$_8$ as a function of temperature
 under different applied fields are presented in Fig. 6. The magnetization shows
 a tendency to decrease at T$\sim$38K. This temperature corresponds to the onset (of
 superconducting transition) T$_c$ measured by ac $\chi$. However at lower temperature
 magnetization starts increasing and goes through a maximum in the ZFC
 measurements and increases monotonically in the FC measurements. The increase
 may be due to the paramagnetic contribution of Gd moments. The magnetization
 curves can be reconciled with as an additive effect due to the contributions to
 the signal of diamagnetism related to superconductivity and the paramagnetic
 response of the Gd$^{3+}$ ions.

\section{Conclusions}

In conclusion, GdSr$_2$Ru$_{1-x}$Cu$_{2+x}$O$_8$(x = 0.05, 0.1, 0.2) exhibits superconducting
 and ferromagnetic transitions, which can be tuned by Cu doping for Ru. T$_c$ is
 increased by partial substitution of Ru by Cu. A doping level of 0.1 gives a
 maximum T$_c$ amongst the investigated compositions. Partial substitution of Ru by
 Cu also has the effect of decreasing the magnetic ordering temperature, from
 139.2K for the undoped composition to 123.2K for 0.2 Cu doped composition.

\end{document}